\documentclass[reprint,nofootinbib,amsmath,amssymb,aps,onecolumn,notitlepage,superscriptaddress]{revtex4-1}

\usepackage[a4paper, margin=2.50cm]{geometry}
\usepackage{graphics}
\usepackage{amsmath}
\usepackage[breaklinks=true,colorlinks]{hyperref}

\newcommand{\nC}{n^{\mathrm{C}}}
\newcommand{\nH}{n^{\mathrm{H}}}
\newcommand{\Dcr}{\Delta_{\mathrm{cr}}}
\newcommand{\Th}{T_{\mathrm{H}}}
\newcommand{\Tc}{T_{\mathrm{C}}}
\newcommand{\Pw}{\mathcal{P}}
\newcommand{\Jh}{J_{\mathrm{H}}}
\newcommand{\Jc}{J_{\mathrm{C}}}
\newcommand{\Qh}{Q_{\mathrm{H}}}
\newcommand{\Tr}{\mathrm{Tr}}
\newcommand{\dd}{\mathrm{d}}
\newcommand{\coloneq}{\mathrel{\mathop:}=}
\newcommand{\eqcolon}{=\mathrel{\mathop:}}
\newcommand{\kB}{k_\mathrm{B}}
\begin{document}

\title{Are quantum thermodynamic machines better than their classical counterparts?}
\author{Arnab Ghosh}
\affiliation{Department of Chemical and Biological Physics, Weizmann Institute of Science, Rehovot 7610001, Israel}

\author{Victor Mukherjee}
\affiliation{Department of Chemical and Biological Physics, Weizmann Institute of Science, Rehovot 7610001, Israel}

\author{Wolfgang Niedenzu}
\affiliation{Institut f\"ur Theoretische Physik, Universit\"at Innsbruck, Technikerstra{\ss}e 21a, A-6020~Innsbruck, Austria}

\author{Gershon Kurizki}
\email{gershon.kurizki@weizmann.ac.il} 
\affiliation{Department of Chemical and Biological Physics, Weizmann Institute of Science, Rehovot 7610001, Israel}

\date{\today}

\begin{abstract}
Interesting effects arise in cyclic machines where both heat and ergotropy transfer take place between the energising bath and the system (the working fluid). Such effects correspond to unconventional decompositions of energy exchange between the bath and the system into heat and work, respectively, resulting in efficiency bounds that may surpass the Carnot efficiency. However, these effects are not directly linked with quantumness, but rather with heat and ergotropy, the likes of which can be realised without resorting to quantum mechanics.
\end{abstract}
\maketitle

%
The term ``quantum thermodynamic machines'' may be understood in two different ways. One is that these are machines ruled by laws that are specific to quantum thermodynamics (QTD), an emerging field that attempts to combine quantum mechanics and thermodynamics \citep{scovil1959three,pusz1978passive,lenard1978thermodynamical,alicki1979quantum,kosloff1984quantum,scully2003extracting,allahverdyan2004maximal,erez2008thermodynamic,delrio2011thermodynamic,horodecki2013fundamental,correa2014quantum,skrzypczyk2014work,brandao2015second,pekola2015towards,uzdin2015equivalence,campisi2016power,rossnagel2016single,kosloff2013quantum,gelbwaser2015thermodynamics,goold2016role,vinjanampathy2016quantum,kosloff2017quantum}. Such laws must depend on quantumness to qualify for QTD.

\par 

The other possible meaning is that these machines are comprised of quantum systems: either all or some of their ingredients are describable quantum-mechanically, but this does 
not imply that these machines function in a quantum fashion. Here we argue, based on our research over the past six years
\citep{gelbwaser2013minimal,gelbwaser2013work,gelbwaser2014heat,gelbwaser2015thermodynamics,niedenzu2016operation,dag2016multiatom,mukherjee2016speed,ghosh2017catalysis,ghosh2018two,niedenzu2018quantum,ghosh2019thermodynamics}, that quantum thermodynamic machines either conform to the latter meaning and do not rely on quantumness \citep{niedenzu2016operation,niedenzu2018quantum} or they are truly quantum, exhibit ``quantum advantage'' \citep{delcampo2014more} but do not contradict the second law of thermodynamics \citep{gelbwaser2013work,ghosh2017catalysis,ghosh2019thermodynamics}.

\par 

\begin{figure}[h]
\begin{center}
\resizebox{0.75\columnwidth}{!}{%
  \includegraphics{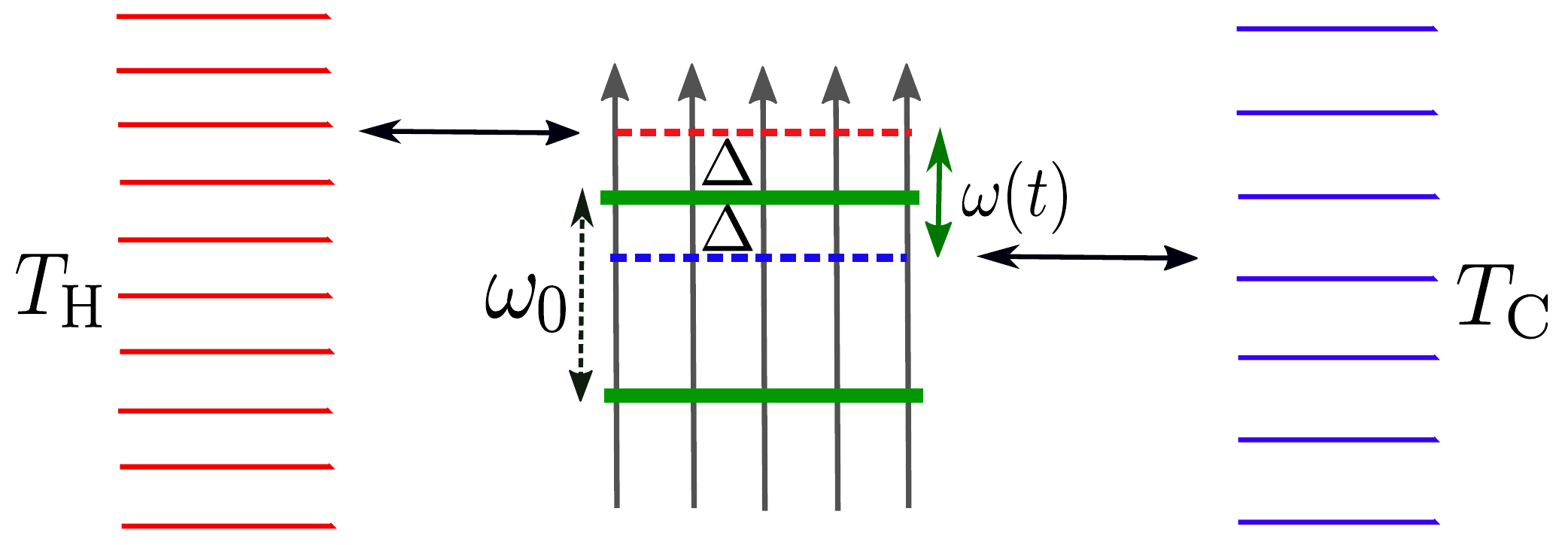} }
\caption{Schematic layout of a minimal heat machine modelled by a two-level working fluid (WF) with resonance frequency $\omega_0$,
driven periodically with time period $2\pi/\Delta$, and exchanging heat with a hot bath at temperature $T_{\rm H}$ and a cold bath at temperature $T_{\rm C}$ \citep{kolar2012quantum,gelbwaser2013minimal}.}
\label{1}       
\end{center}
\end{figure}
The first machine we analysed \citep{gelbwaser2013minimal} was deemed to be the minimal or simplest heat machine based on a quantum system --- a qubit. 
The qubit with resonance frequency $\omega_0$ is the working fluid (WF) of the machine. It is permanently coupled to cold and hot thermal baths with different, 
non-overlapping spectra. The qubit is driven periodically by a classical field which acts as a piston that causes periodic modulation of the qubit frequency $\omega(t)$ (Fig.~\ref{1}). 
The modulation period $2\pi/\Delta$ constitutes the machine cycle time. The merit of this model is that it is amenable to a complete analysis within the weak-coupling, Markovian 
approximation for the system-bath dynamics \citep{gelbwaser2013minimal,gelbwaser2015thermodynamics}.

\par 

This analysis yields the result that the machine may act as a refrigerator (or heat pump) under the condition
\begin{eqnarray}\label{heat-pump-cond}
\nC(\omega_0-\Delta) > \nH(\omega_0+\Delta),
\end{eqnarray}
and as a heat engine under the converse condition
\begin{eqnarray}\label{heat-engine-cond}
\nC(\omega_0-\Delta) < \nH(\omega_0+\Delta).
\end{eqnarray}
Here $\nC(\omega_0-\Delta)$ and $\nH(\omega_0+\Delta)$ are the cold and hot bath thermal occupancies at the downshifted and upshifted transition frequencies, respectively.
These conditions characterise the optimal scenario wherein the qubit at the upshifted frequency only couples to the hot bath and at the downshifted frequency to the cold bath.  
\par 

Equivalently to Eqs.~\eqref{heat-pump-cond}-\eqref{heat-engine-cond}, the machine acts as a heat engine whose piston extracts power ($\Pw<0$) provided that the (positive)
modulation frequency $\Delta$ is bounded (from above) by 
\begin{eqnarray}\label{delta-<-delta-cr}
\Dcr=\omega_0\frac{\Th-\Tc}{\Th+\Tc},
\end{eqnarray}
$\Th$ and $\Tc$ being the hot and cold bath temperatures, respectively. The efficiency, defined as the ratio of the extracted power $-\Pw$ to the input heat input current $\Jh$ from the hot bath, 
grows with $\Delta$ until the Carnot bound is attained at $\Dcr$,
\begin{eqnarray}\label{carnot-bound}
\eta=\frac{-\Pw}{\Jh}=\frac{2\Delta}{\omega_0+\Delta}\leq 1-\frac{\Tc}{\Th}.
\end{eqnarray}

\par 

As the modulation frequency exceeds the critical value, i.e., $\Delta > \Dcr$, the machine becomes a refrigerator for the cold bath. It consumes power ($\Pw>0$) from the piston and converts it into cold current $\Jc$ as characterised by the coefficient of performance (COP) that reaches its maximal value at $\Delta=\Dcr$,
\begin{eqnarray}\label{cop-bound}
\mathrm{COP}=\frac{\Jc}{\Pw}=\frac{\omega_0-\Delta}{2\Delta} \leq \frac{\Tc}{\Th-\Tc}.
\end{eqnarray}

\par 
These lucid, simple results show clearly that although the WF is a qubit, there is nothing uniquely quantum-mechanical about the machine performance, which 
adheres to the standard thermodynamic bound.

\par 
\begin{figure}[h]
\begin{center}
\resizebox{0.4\columnwidth}{!}{%
  \includegraphics{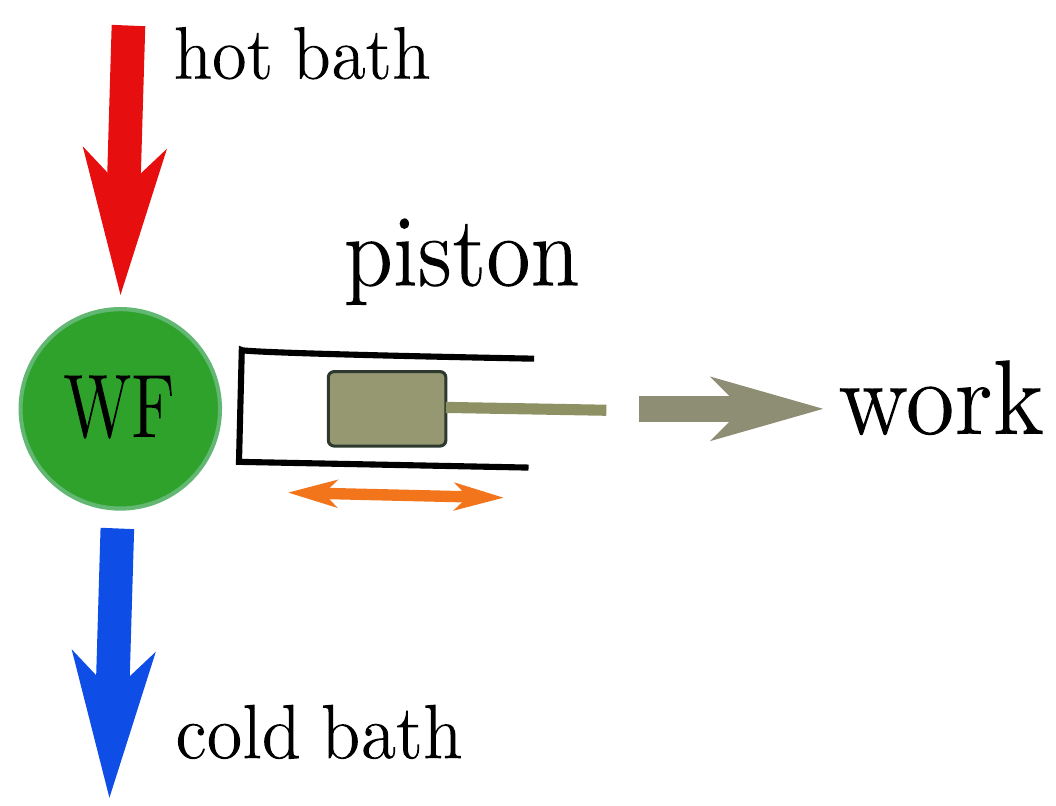} }
\caption{Universal heat-machine cycle modelled by a WF energized by a hot bath, and dumping heat in a cold bath. The WF is driven cyclically by a classical field which acts as a piston that extracts work.}
\label{3}       
\end{center}
\end{figure}
Yet, the field of quantum-thermodynamic machines has been propelled by ingenious porposals to benefit from quantum resources embodied by non-thermal baths
\citep{scully2003extracting,dillenschneider2009energetics,huang2012effects,abah2014efficiency,rossnagel2014nanoscale,niedenzu2016operation,manzano2016entropy,hardal2015superradiant,klaers2017squeezed,agarwalla2017quantum,niedenzu2018quantum}. 
Schematically, such machines have the same ingredients as conventional Carnot heat engines (Fig.~\ref{3}). However, at least the hot bath, which is the source of energy, may have non-thermal properties that stem from its quantum-mechanical preparation. The question has been posed whether a cycle energised by such a bath must abide by the Carnot efficiency bound derived in 1824 for steam engines \citep{carnotbook}, 
\begin{eqnarray}\label{carnot-bound-1824}
\eta=\frac{-W}{\Qh}\leq 1-\frac{\Tc}{\Th}=:\eta_\mathrm{C},
\end{eqnarray}
where the efficiency is the ratio of the work output $-W$ to the heat input $\Qh$. Two crucial assumptions have been made in Eq.~\eqref{carnot-bound-1824}: (i) that the input from the ``hot'' 
(better: energising) bath is indeed heat, and (ii) that this bath has a ``temperature'' $\Th$, although a non-thermal bath need not have one.

\par 
\begin{figure}[h]
\begin{center}
\resizebox{0.75\columnwidth}{!}{%
  \includegraphics{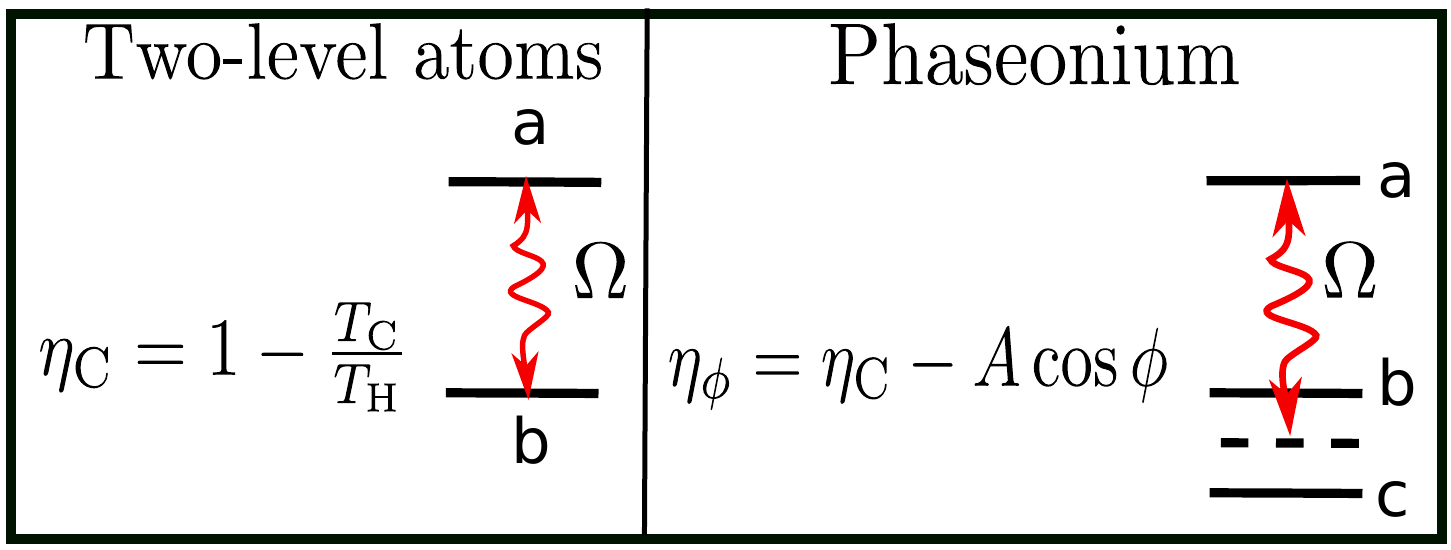} }
\caption{Left -- the Carnot bound $\eta_\mathrm{C}$ for a micromaser cycle fuelled by a bath of two-level atoms. Right -- the same for a bath of three level ``phaseonium'' atoms. Here $A$ and $\phi$ are the modulus and phase of the coherence induced between the nearly-degenerate levels b and c. This bound may exceed $\eta_\mathrm{C}$ \citep{scully2003extracting}.}
\label{phaseonium}       
\end{center}
\end{figure}
Before addressing these issues, we consider the specific setups which have promoted our general investigation of these issues \citep{niedenzu2016operation,niedenzu2018quantum}.
The first setup, whose study by Scully et. al. \citep{scully2003extracting} pioneered the field, consists of an engine that is energised by ``phaseonium'' fuel. The latter are three-level atoms whose lower two near-degenerate 
levels are coherently superimposed with a phase $\phi$ (Fig.~\ref{phaseonium}). The consecutive interactions of these atoms with the WF (a single cavity-field mode) conform to the  micromaser
model whereby the bath \emph{appears} to be at temperature $\Th(\phi)$, which is now $\phi$-dependent. Consequently, whenever a $\phi$ is chosen such that $\Th(\phi)$
exceeds $\Th$ in the absence of coherence, then the resulting Carnot bound becomes higher than the standard (incoherent) Carnot bound. Is this a quantum advantage? Not from the point of view of the
WF (cavity field mode) that interacts with a bath at a temperature $\Th(\phi)$ --- the WF has no other bath except the phaseonium.

\par 
We next turn to another setup that was proposed by Ro\ss{}nagel \textit{et al.} \citep{rossnagel2014nanoscale}: an Otto cycle that is energised by a squeezed-thermal bath. In this cycle the strokes are realized by adiabatic compression and expansion of the WF, its consecutive coupling to and decoupling from the cold and the hot baths, the only difference from the standard cycle being that the hot bath is not in a thermal state (at temperature $T_\mathrm{H}$) but rather in a squeezed-thermal state (described by the temperature $T_\mathrm{H}$ and the squeezing parameter $r$). 

\par 

How does the squeezing affect the cycle efficiency? The tendency of most works on the subject \citep{huang2012effects,abah2014efficiency,rossnagel2014nanoscale,manzano2016entropy,klaers2017squeezed} has been to identify the entire energy delivered by the hot bath to the WF as heat. If we attribute to this definition and the extra (squeezing) energy to an effective temperature $\Th(r)$ that increases with the squeezing parameter, then we may again deduce from this analysis that the cycle may surpass the standard Carnot bound provided $\Th(r)$ is higher than $\Th$ without squeezing \citep{rossnagel2014nanoscale,manzano2016entropy}. 

\par 

However, there is a missing piece in this puzzle: can we really claim that heat and squeezing are interchangeable resources? 

\par 

To answer this question, we have to properly unravel the energy exchange between a quantum system and a quantum bath, namely, reach better understanding of the \textit{first law of thermodynamics} in the quantum domain. To this end, we start from Alicki's pioneering decomposition \citep{alicki1979quantum} of the change in the mean energy of a quantum system that is both driven by a time-dependent system Hamiltonian $H(t)$ and is coupled to a quantum bath. Such a change in the mean system energy 
\begin{eqnarray}\label{mean-energy}
E(t)=\Tr[\rho(t)H(t)]
\end{eqnarray}
for the reduced system density operator $\rho(t)$ was decomposed by Alicki as
\begin{subequations}\label{eq_defs_Ediss_work} 
\begin{equation}\label{eq_first_law}
  \Delta E(t)=\mathcal{E}(t)+W(t), 
\end{equation}
and split into heat
\begin{equation}\label{eq_def_DeltaEdiss}
    \mathcal{E}(t)\coloneq\int_0^t\Tr[\dot\rho(t^\prime)H(t^\prime)]\dd t^\prime
  \end{equation}
and work
\begin{equation}\label{eq_def_work}
    W(t)\coloneq\int_0^t\Tr[\rho(t^\prime)\dot H(t^\prime)]\dd t^\prime.
  \end{equation}
\end{subequations} 
Here $\rho(t)$ is the reduced density operator of the system obtained by tracing out the bath degrees of freedom and $H(t)$ is the controlled Hamiltonian for the system. In Refs.~\citep{pusz1978passive,lenard1978thermodynamical}, Eq.~\eqref{eq_def_work} is interpreted as work because it is associated with a change $\dot H(t)$ in the driving Hamiltonian. Alternative definitions of quantum work in the literatures include, for example, the concept of ``work operator'' in the context of quantum measurement processes~\citep{talkner2016aspects,baeumler2018fluctuating,PhysRevE.75.050102,Hanggi2015,RevModPhys.83.771,
RevModPhys.83.1653,PhysRevLett.102.210401,PhysRevLett.118.070601,PhysRevLett.93.048302,PhysRevE.71.066102,PhysRevE.78.011116}.

\par 

Here, we are interested in the question whether \eqref{eq_def_DeltaEdiss} necessarily correspond to entropy change (in the regime of weak coupling between the system and the baths, neglecting correlations between the system and the baths~\cite{breuerbook}). On the face of it, it does, because it arises from a change in the state of the system $\dot\rho(t)$ that may contribute to entropy change. However, as shown below, this is not always the case. In fact, there are isentropic processes associated with $\dot{\rho}(t)\neq 0$, which physically correspond to work rather than heat exchange.

\par 
\begin{figure}[h]
\begin{center}
\resizebox{0.75\columnwidth}{!}{%
  \includegraphics{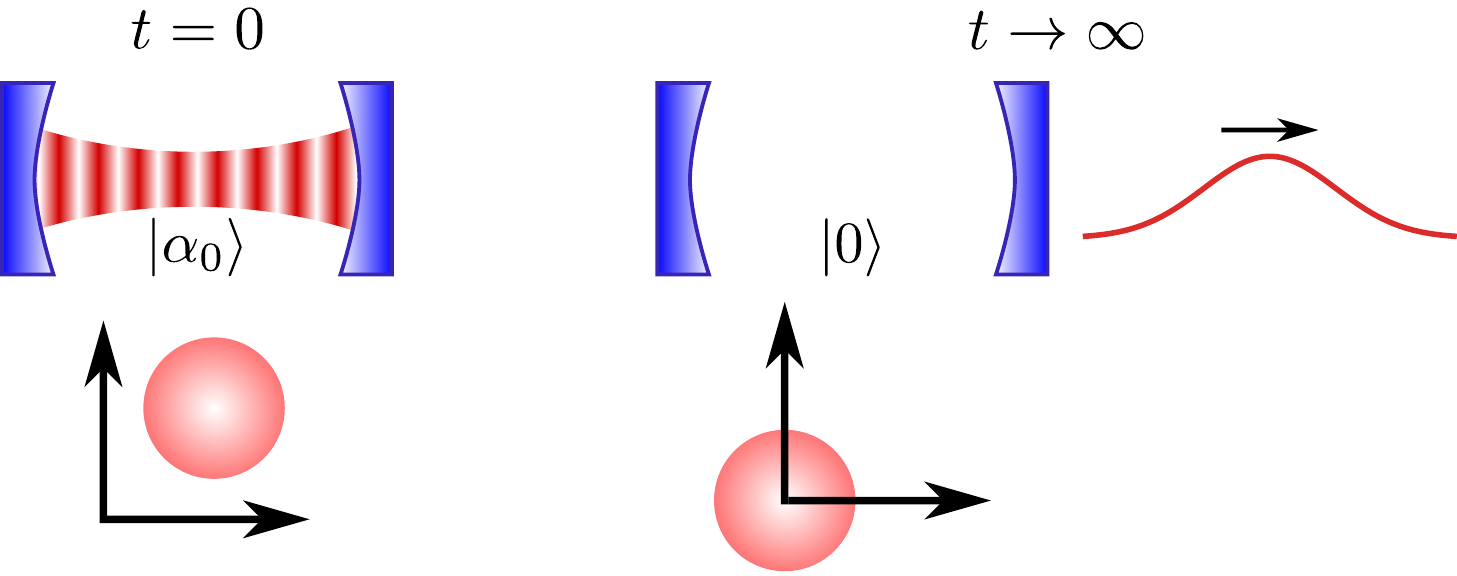} }
\caption{Energy transfer from a relaxing cavity mode initially prepared in a coherent state $|\alpha_0\rangle$ cannot be associated with heat transfer, because it is isentropic. Figure adapted from \citep{niedenzu2018quantum}.}
\label{4}       
\end{center}
\end{figure}
As a simple example of such a process, which may constitute part of a cycle, consider a single-mode cavity field initially prepared in a coherent state $|\alpha_0\rangle$. It leaks out of the cavity through the front mirror, until at long times the cavity-field state becomes the vacuum $|0\rangle$ (Fig.~\ref{4}). The point is that the state has changed and so has its mean energy $E(t)$ but not its purity or entropy. This is an example of a system that is transformed from a \textit{non-passive} state, here a coherent state with non-zero amplitude $|\alpha(t) \neq 0\rangle$, to a passive state, here the vacuum state $|0\rangle$. A passive state is a state that does not allow to extract work from the system under cyclic unitary transformations \citep{pusz1978passive,lenard1978thermodynamical}. As long as the Hamiltonian is non-degenerate, there is a unique passive state for each non-passive state: the two are unitarily related. Thus, the leakage of the field from the cavity is a change in the \textit{degree of non-passivity} but not in entropy. Such a change has been dubbed \textit{ergotropy} change \citep{allahverdyan2004maximal}. We note that ergotropy cannot be readily measured, as \textit{any} measurement would cause back-action which is typically not unitary~\cite{talkner2016aspects}. It would thus be interesting to study the issue of back-action on ergotropy~\citep{venkatesh2015quantum}.

\par 

\begin{figure}[h]
\begin{center}
\resizebox{0.75\columnwidth}{!}{%
  \includegraphics{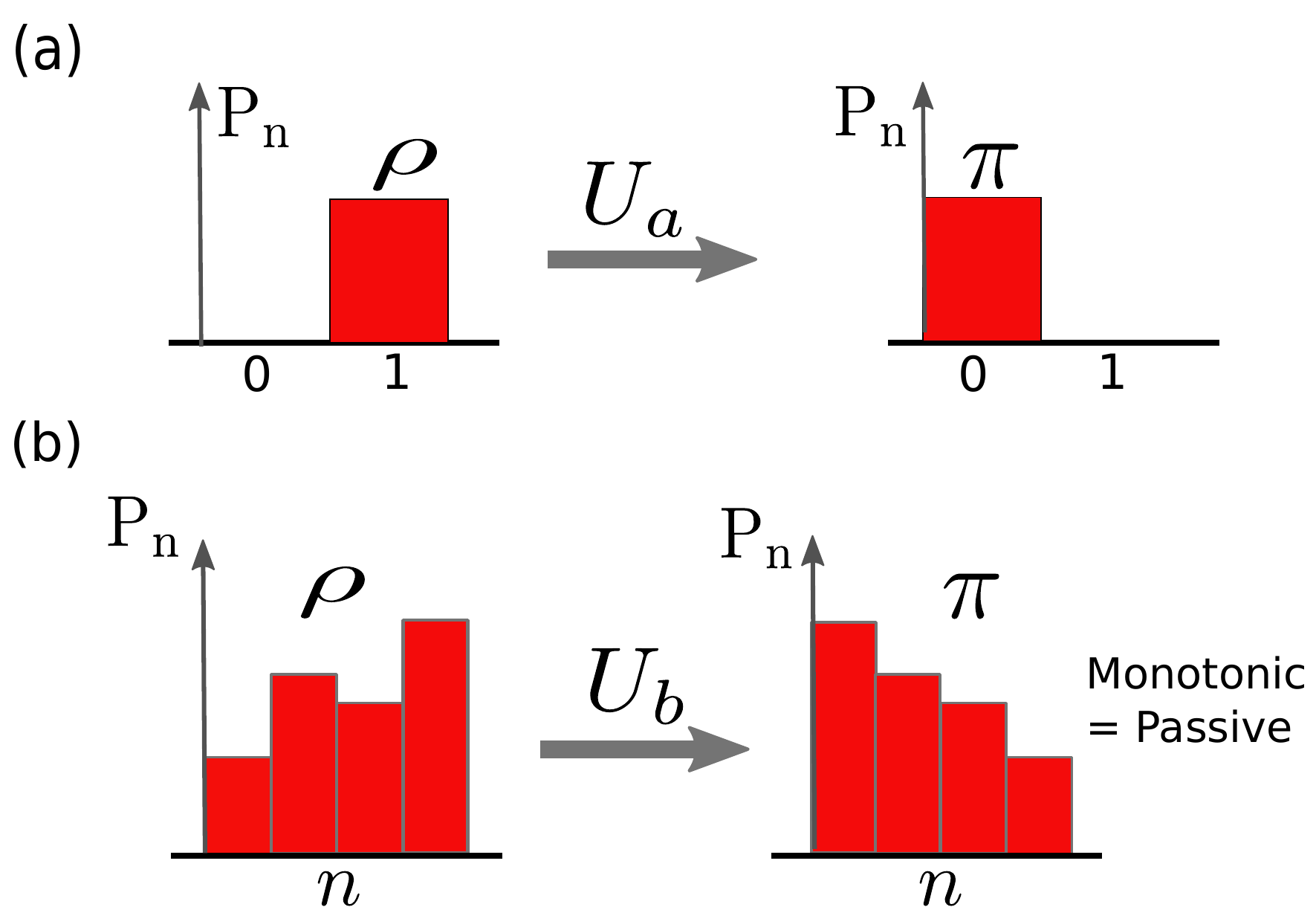} }
\caption{Transformation of an initial non-passive state $\rho$ to its passive state $\pi$ via a unitary reshuffeling of the eigenvalues. (a) Ergotropy release by a unitary transformation from Fock state $|1\rangle$ to $|0\rangle$, the latter being passive. (b) The same for an arbitrary initial mixture of Fock states.}
\label{5}       
\end{center}
\end{figure}

Passivity of a quantum state implies a monotonically-decreasing distribution of energy  eigenvalues. For details please see Refs.~\citep{pusz1978passive,lenard1978thermodynamical}. As shown in Fig.~\ref{5}, a unitary transformation from a Fock state $|n\neq 0\rangle$ to the vacuum $|0\rangle$ (Fig.~\ref{5}a) or from a non-monotonic distribution of Fock states to a reshuffled monotonic distribution with the same Fock-state ingredients (Fig.~\ref{5}b) releases all the ergotropy of the initial non-passive state $\rho$, resulting in the passive state $\pi$. Being isentropic, it should be distinguishable from dissipative change in passive energy. 
\par 

Such distinction is effected by the decomposition of the energy exchanged between the system and the bath into 
\begin{subequations}\label{eq_defs_heat_passive_work}
\begin{equation}\label{eq_DeltaEdiss_decomposition}
  \mathcal{E}(t)=\int_0^t\Tr[\dot\rho(t^\prime)H(t^\prime)]\dd t^\prime
  =\mathcal{Q}(t)+\Delta\mathcal{W}|_\mathrm{diss}(t), 
\end{equation}
where the non-unitary changes in passive energy and system ergotropy are given by \citep{niedenzu2018quantum}
  \begin{equation}\label{eq_def_heat}
    \mathcal{Q}(t)\coloneq\int_{0}^{t} \Tr[\dot{\pi}(t^\prime)H(t^\prime)]\dd t^\prime
  \end{equation}
and
\begin{equation}\label{eq_def_DeltaW_diss}
    \Delta\mathcal{W}|_\mathrm{diss}(t)\coloneq\int_0^t\Tr\Big[\big(\dot\rho(t^\prime)-\dot\pi(t^\prime)\big)H(t^\prime)\Big]\dd t^\prime,
  \end{equation}
\end{subequations}
respectively. The instantaneous passive state $\pi(t)$ and its derivative $\dot{\pi}(t)$ are obtained by unitary transformations from the time evolution of $\rho(t)$. Hence, Eq.~\eqref{eq_def_heat} is always associated with a change in the von~Neumann entropy $\mathcal{S}(\rho(t))=-\kB\Tr[\rho(t)\ln \rho(t)]$ because $\mathcal{S}$ of a non-passive state $\rho(t)$ is same as that of its unitarily transformed passive state $\pi(t)$. In accordance with the existing  literature, we consider the von-Neumann entropy to be relevant in thermodynamic settings, whether in or out of equilibrium~\citep{spohn1978entropy,alicki1979quantum,kosloff2013quantum,binder2019thermodynamics,vinjanampathy2016quantum}. Thus, in analogy to thermodynamics, we hereafter refer to Eq.~\eqref{eq_def_heat} as \textit{heat} transfer because of its entropy-changing character.

\par 

The above decomposition of the energy exchange with the bath into heat and ergotropy exchange has led us to suggest the following inequalities for the entropy change over long times $t \rightarrow \infty$ of the \textit{system} (assuming the bath is \textit{thermal}, i.e., its temperature $T$ to be immutable) \citep{niedenzu2018quantum,ghosh2019thermodynamics},
\begin{subequations}\label{our-inequality-nat-comm}
\begin{equation}\label{eq_DeltaS_QdT}
  \Delta\mathcal{S}\geq\frac{\mathcal{Q}}{T}
\end{equation}
for a system governed by a constant Hamiltonian or
\begin{equation}\label{eq_DeltaS_Qth}
  \Delta\mathcal{S}\geq\frac{\mathcal{E}^\prime}{T},
\end{equation}
\end{subequations}
for a system driven by a time-dependent Hamiltonian, where $\mathcal{E}^\prime$ is the energy which would be exchanged with the bath under such driving if the initial state were the passive counterpart $\pi(0)$ of the actual initial state $\rho(0)$. It is clear, particularly from \eqref{eq_DeltaS_QdT}, that $\Delta \mathcal{S}$ is here bounded from below only by heat exchange.

\par

Our suggested inequality \eqref{our-inequality-nat-comm} must be contrasted with Spohn's \citep{spohn1978entropy}
\begin{equation}\label{eq_DeltaS_QdT_DeltaW}
  \Delta\mathcal{S}\geq\frac{\mathcal{E}}{T}=\frac{\mathcal{Q}+\left.\Delta\mathcal{W}
  \right|_\mathrm{diss}}{T},
\end{equation}
where $\Delta \mathcal{S}$ is bounded from below by the sum of heat and ergotropy exchange. Although inequality \eqref{eq_DeltaS_QdT_DeltaW} is as much consistent with the second law (the Clausius inequality \citep{clausiusbook1}) as our inequality \eqref{our-inequality-nat-comm},  the latter is much tighter than the former, because, for an initial non-passive state, $\Delta\mathcal{W}|_\mathrm{diss} \leq 0$ in a relaxation process of the system towards its thermal steady-state. 

\par 
\begin{figure}[h]
\begin{center}
\resizebox{0.75\columnwidth}{!}{%
  \includegraphics{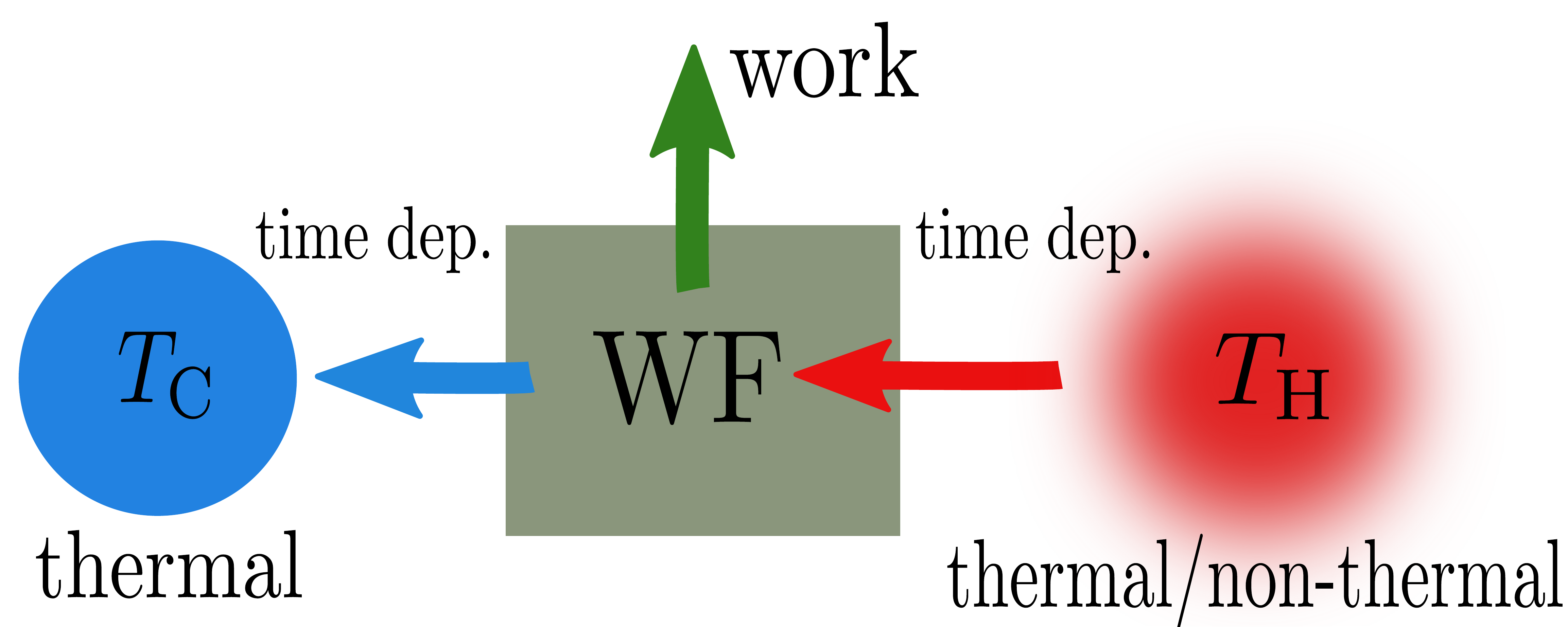} }
\caption{Universal layout of a cyclic machine powered by either a thermal or non-thermal (e.q. squeezed) bath.}
\label{6}       
\end{center}
\end{figure}
These considerations have been used by us to derive a generalised efficiency bound for cycles in which the WF may be energised by either a thermal or a non-thermal bath \citep{niedenzu2016operation,niedenzu2018quantum}. As shown schematically in Fig.~\ref{6}, such a cycle consists of time-dependent periodic change of the coupling between the WF and the energising (thermal or non-thermal) bath (right ellipse) and entropy dumping into a cold thermal bath (left circle). 

\par 

Our general efficiency bound \citep{niedenzu2018quantum} is 
\begin{equation}\label{eq_etamax_gen}
  \eta\leq 1-\frac{\Tc}{\Th}\frac{\mathcal{E}^\prime_\mathrm{h}}{\mathcal{E}_\mathrm{h}}\eqcolon\eta_\mathrm{max} \le 1.
\end{equation}
On the r.h.s, the temperature ratio of the two baths is multiplied by the ratio of the heat exchange $\mathcal{E}^\prime_\mathrm{h}$ to the total energy exchange $\mathcal{E}_\mathrm{h}$, which combines both heat and ergotropy exchange. Here $\mathcal{E}^\prime_\mathrm{h}$ is the heat that the WF would have obtained in the same cycle if the hot bath was thermal. It then follows that in the absence of ergotropy transfer from the bath to the WF, as for a thermal bath, the Carnot bound is recovered. By contrast, if the heat exchange with the energised bath is much less than total energy exchange, because ergotropy transfer dominates, the efficiency bound approaches $1$ and surpasses the Carnot efficiency. This, however, should in no way imply a \textit{surpassing} of the Carnot bound: it means that the Carnot bound is inapplicable to such a scenario, wherein ergotropy rather than heat is exchanged with the bath. In fact, the second law, from which the Carnot bound follows, only relates to entropy-changing processes, whereas ergotropy exchange is isentropic. Here we have not considered the cost of producing the state of the bath. By the same token, it is customary not to consider the cost of producing thermal baths for conventional heat engines~\cite{callenbook}.

\par 

Can the cycle described above, wherein ergotropy and non-passivity play a central role, be deemed genuinely quantum, or at least exhibit quantum advantage? Not necessarily, since the energising bath in question is in a squeezed-thermal state and such a state exhibits genuine quantumness only if the temperature $\Th$ associated with its thermal component is low enough, such that the state approximates the squeezed-vacuum state \citep{niedenzu2016operation,kim1989properties}.

Although we have explicitly considered above ``semiclassical'' engines whose cycle is effected by a classical periodic driving field (``piston''), similar conclusions hold for fully quantised machines, wherein the quantum state of the piston is explicitly accounted for~\cite{gelbwaser2014heat,ghosh2017catalysis,ghosh2018two}: In the latter class of machines it is the ergotropy (non-passivity) of the piston state and its heating (thermalisation), which are not exclusively related to quantumness, that determine the efficiency and power output of the machine.

Since our discussion has revolved around the decomposition of system-bath energy exchange into work and heat transfer, it should be stressed that we cannot identify any alternative, physical, quantifiers of these processes. Namely, we find the physical arguments in favour of the uniqueness of the present quantifiers to be compelling.

To sum up, interesting effects arise when one considers cyclic machines where both heat and ergotropy transfer take place. Such effects correspond to unconventional decompositions of energy exchange between the bath and the WF into heat and work, resulting in efficiency bounds that may exceed Carnot's. This is allowed because the thermodynamic Carnot bound is only valid for machines energised by heat transfer. However, these effects are not directly linked with quantumness, but rather with heat and ergotropy exchange, the likes of which can be constructed without resorting to quantum mechanics. 

These conclusions have to be revised when multiple quantum-correlated (e.g. entangled) machines are compared to their classical counterparts~\citep{niedenzu2018cooperative}.

\textbf{Acknowledgements:}
We acknowledge the support of the DFG, ISF and SAERI (G.K.), VATAT (V.M.) and the ESQ fellowship of the Austrian Academy of Sciences (\"OAW) (W.N.).

\textbf{Author contribution statement:} Arnab Ghosh, Victor Mukherjee and Wolfgang Niedenzu (listed alphabetically) have contributed to the research results and physical insights involved in this paper. Gershon Kurizki has contributed to the formulation of the overall physical picture and has mostly written the article.

\end{document}